\documentclass[amsmath,superscriptaddress,showpacs]{revtex4}

\usepackage{amsfonts}
\usepackage{amsmath}
\usepackage{amssymb}
\usepackage{graphicx}
\usepackage{txfonts}

\renewcommand{\vec}[1]{\mathbf{#1}}
\newcommand{\tx}{{\tilde{x}}}

\begin{document}
	\title{$\mathbb{Z}_4$ Parafermions \& the $8\pi$-periodic Josephson effect in interacting Rashba nanowires}
	\author{Chris J. Pedder}
	\affiliation{Physics and Materials Science Research Unit, University of Luxembourg, L-1511 Luxembourg}
	\author{Tobias Meng}
	\affiliation{Institut f\"ur Theoretische Physik, Technische Universit\"at Dresden, 01062 Dresden, Germany}
	\author{Rakesh P. Tiwari}
	\affiliation{Department of Physics, University of Basel, Klingelbergstrasse 82, CH-4056 Basel, Switzerland}
	\author{Thomas L. Schmidt}
	\email{thomas.schmidt@uni.lu}
	\affiliation{Physics and Materials Science Research Unit, University of Luxembourg, L-1511 Luxembourg}
	
	\date{\today}

	\begin{abstract}
    We demonstrate the appearance of zero-energy bound states satisfying the $\mathbb{Z}_4$ parafermionic algebra in interacting quantum wires with Rashba spin-orbit coupling and proximity-induced superconductivity.
    The fourfold degeneracy of these states is protected by time-reversal symmetry and fermion-parity conservation, and leads to an $8 \pi$ periodicity of the Josephson current due to the tunneling of fractionalized excitations with charge $e/2$. Even in the presence of perturbations, we propose that this periodicity will remain visible in driven, current-biased Shapiro step measurements on current state-of-the-art Rashba wires.
	\end{abstract}
	
	
	\maketitle
	
\section{Introduction.}
One of the most active fields in physics in recent years has been the search for Majorana fermions \cite{Nayak+2008,leijnse12,Alicea+2012,beenakker13}. Despite being theoretically predicted more than 70 years ago \cite{majorana37}, the existence of Majorana fermions as \emph{elementary} particles is still unclear. However, it was recently recognized that they can occur as quasiparticles in various solid-state systems \cite{read00,Ivanov+2001,Kitaev+2001,Fu+2008,Fu+2009,Sato+2009,Sato+2009B,Oreg+2010,Lutchyn+2010}. For instance, certain one-dimensional (1D) wires are believed to host localized Majorana bound states (MBS) due to the combined effects of Rashba spin-orbit coupling (RSOC), proximity-coupling to a superconductor, and an externally applied magnetic field \cite{Oreg+2010,Lutchyn+2010}. These MBS can be understood using simple single-particle quantum mechanics. Exotic and closely related bound states, called parafermions, have been predicted to descend from strongly correlated phases such as the edges of fractional quantum Hall systems \cite{fendley12,Lindner+2012,Clarke+2013,alicea15}.

The practical interest in these unusual quasiparticles stems primarily from their suitability for topologically-protected quantum computation \cite{Nayak+2008}. Exploiting their non-Abelian exchange statistics, many logic gates can be implemented by braiding, an operation which is robust to local perturbations. Indeed, the prospect of intrinsically decoherence-free qubit operations is a strong driving force for this research field, and has the potential to revolutionize quantum computation.

If experimentally confirmed, MBS and parafermions will be the first discovered particles with non-Abelian exchange statistics. Edge states of two-dimensional topological insulators, ferromagnetic chains on superconductors, as well as semiconductor nanowires with RSOC have recently shown experimental signatures that are consistent with the appearance of MBS \cite{Deng+2012,Mourik+2012,Wesperen+2013,Nadj+2014}. In particular, recent work \cite{rokhinson12,Maier+2015,Wiedenmann+2015} suggests that it is possible to observe $4\pi$ periodic components of the Josephson current arising from Majorana fermions, giving a window on their twofold degeneracy. Measured current-voltage curves for proximitized edge states of HgTe and nanowires with RSOC have shown the disappearance of odd numbered ``Shapiro steps'' consistent with theoretical predictions \cite{Dominguez+2012}. In contrast, experimental results indicating the presence of parafermions do not yet exist, largely because the proposed experimental setups are difficult to realize.

In this paper, we aim to tackle this lack of experimental realization of parafermions by describing a rather simple arrangement in which a single interacting 1D Rashba wire exhibits fourfold degenerate zero-energy states obeying a $\mathbb{Z}_4$ parafermionic algebra (see Refs.~\cite{Zhang+2014,klinovaja13b,Klinovaja+2015,Orth+2015,Haim+2014} and references therein). Our proposed system exploits a generic interaction process which has so far not been examined in quantum wires, called spin-umklapp scattering, whereby two spin-up fermions are scattered into two spin-down fermions. Such an interaction arises from virtual transitions between subbands, and for a suitable choice of the chemical potential opens a partial gap in the energy spectrum of the wire. Adding proximity-induced superconductivity then leads to the emergence of parafermion bound states at the ends of the wire, whose fourfold degeneracy is protected by Kramers' theorem as long as time-reversal symmetry (TRS) is not broken. As a definitive experimental signature, we point out that tunneling of the fractionalized quasiparticles with charge $e/2$ will result in an $8 \pi$ periodic component of the Josephson current, which could be seen in Shapiro step experiments. We note that the required interaction strengths have already been reached in quantum wires \cite{Hevroni+2015}, and that the opening of the spin-umklapp gap may already have been seen in recent experiments (e.g. \cite{Zumbuhl+2014}). Moreover, the predicted fourfold degeneracy and the associated Josephson effect are robust to weak disorder, of the magnitude of the gap which is opened by spin-umklapp scattering.
	
In section 2, we introduce the model for an interacting Rashba wire and derive the spin-umklapp scattering term. We then bosonize this model in section 3, and present a renormalization group (RG) analysis for the flow of the system parameters. We find there exist regimes in which fourfold degenerate edge states can be found. In section 4, we describe the implications of the degenerate ground state for Josephson effect measurements, and suggest a definitive Shapiro step measurement to identify the $8\pi$ contribution to the Josephson current arising from parafermions, even when this contribution is sub-dominant. Finally, in section 5, we discuss the relevance of our work for experimental investigations.
	
\section{Model.}
We consider a long quasi-1D nanowire along the $x$ direction which is harmonically confined in the $y$ and $z$ directions. The interplay of the intrinsic spin-orbit coupling of a material and the breaking of inversion symmetry of a particular geometric arrangement, due to either the presence of a substrate, or to the application of an out of plane electric field, gives rise to RSOC \cite{winkler_book}. In the latter case, the strength of the electric field may be used to tune the magnitude of the RSOC, denoted as $\alpha_R$. The dynamics in the $z$-direction is not affected by the Rashba coupling and can be safely ignored. Thus, we can model a finite-width wire by using the following 2D Hamiltonian including RSOC with strength $\alpha_R$ \cite{moroz00,moroz00a,Starykh+2008}
\begin{equation}
H = \frac{ p_x^2+p_y^2}{2m} + \frac{1}{2} m \omega^2 y^2 + \alpha_R (\sigma_x p_y - \sigma_y p_x),
\label{RashbaHamiltonian}
\end{equation}
where $p_x$ and $p_y$ are the momentum components in the $x$- and $y$-directions, and $\sigma_{x,y}$ are Pauli matrices. The transversal confinement is modelled as a harmonic potential with frequency $\omega$. The system has translational invariance along the $x$-direction but is strongly confined along the $y$-direction, which leads to the appearance of higher excited bands separated from the lowest band by a spacing determined by the inverse width of the wire. Introducing raising and lowering operators, $a^\dag$ and $a$, one finds that $H = H_0 + H_1$, where
\begin{eqnarray}
H_0 &= \omega \left( a^\dagger a +\frac{1}{2} \right) +\frac{p_x^2}{2m} - \alpha_R \sigma_y p_x, \\
H_1 &= ig \sigma_x (a^\dagger - a),
\end{eqnarray}
and $g = \alpha_R \sqrt{m \omega/2}$. Since eigenstates of $H_0$ have a quantized spin in $\sigma_y$ direction, the form of $H_1$ ensures that transitions between neighboring subbands, which are caused by $a$ and $a^\dag$, always involve a spin flip. We will account for the possibility of virtual transitions between the lowest subband and the first excited subband with energy $\omega$ by integrating out the coupling $H_1$ between the subbands. We do this by means of a Schrieffer-Wolff transformation, $H^\prime = e^{-S} H e^S$ where $[S,H_0] = H_1$. Up to second order in $m\alpha_R^2/\omega$ one finds
\begin{equation}
H^\prime = H_0 - \frac{1}{2} [S,H_1] = \frac{p_x^2}{2m} - \alpha_R \sigma_y p_x,
\end{equation}
which tells us that it would be consistent to simply ignore the dynamics in $y$-direction in Eq.~(\ref{RashbaHamiltonian}) \cite{Starykh+2008}. Deviation of the spectra from the parabolic form occurs only at order $(m \alpha_R^2/\omega)^3$ \cite{moroz00,moroz00a}, and is not relevant for our current discussion. Note, however, that the associated change in Fermi velocities can have interesting effects for weak interactions \cite{kainaris15}.
	
\begin{figure}[t]
	\centering
	\includegraphics{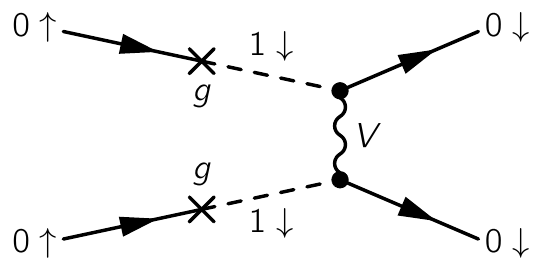}
	\caption{A process in which virtual transitions ($g$) between the lowest and the first excited subbands of the quantum wire allow for a spin-non-conserving umklapp process to be generated from a spin conserving interaction $V$.}
	\label{VirtualUmklapp}
\end{figure}

In the interacting case, the interaction term also gets modified by the Schrieffer-Wolff transformation. A generic density-density interaction potential $V(\vec{r}_1 - \vec{r}_2)$ where $\vec{r}_i = (x_i, y_i)$, clearly conserves spin. However, since virtual transitions to higher bands come with a spin-flip and the interaction potential $V(\vec{r})$ mixes states in different subbands, spin-umklapp scattering is generated as shown in Fig. \ref{VirtualUmklapp}.

After the Schrieffer-Wolff transformation, the interaction potential for the lowest subband takes the form $V^\prime = e^{-S} V e^{S} = V_\rho + V_{\rm{sf}} + V_{\rm{sx}}$, with a density-density interaction $V_\rho$, a spin-flip term $V_{\rm{sf}}$ which changes the total spin of the interacting particles, and a spin-exchange term $V_{\rm{sx}}$, which allows particles in the lowest band to exchange their spins, but conserves the total spin.

\begin{figure}[t]
	\centering
	\includegraphics[width=0.45\linewidth]{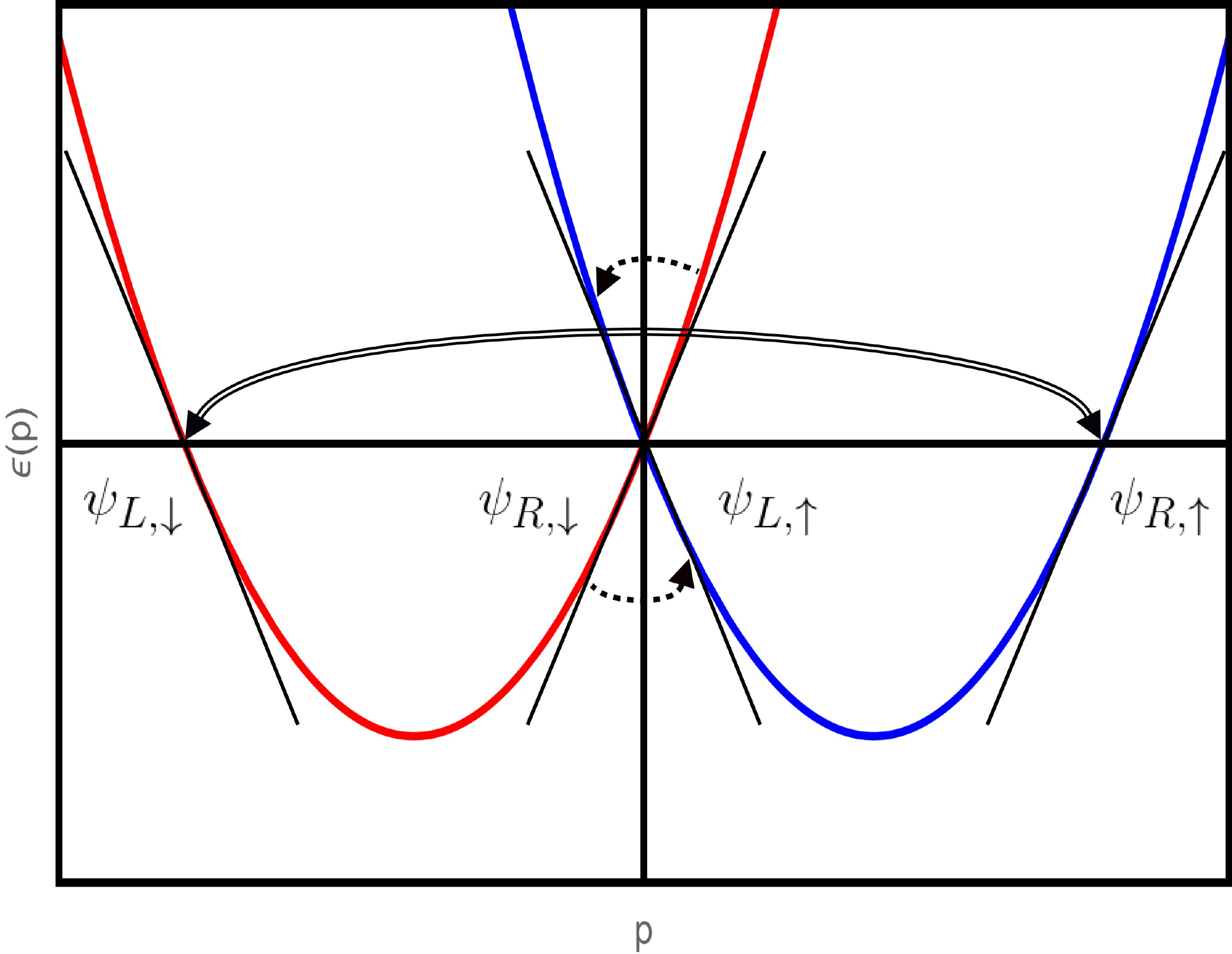}
	\caption{(color online) This figure shows the position of the chemical potential and the four low-energy linearized modes used to bosonize the system. Arrows indicate the RG-relevant interaction processes between these modes: umklapp scattering between the two modes near $k=0$ (dotted lines) and proximity-induced $s$-wave superconductivity, which pairs modes of opposite physical spin (double line).}
	\label{Interactions}
\end{figure}

Placing the chemical potential at the Dirac point, as shown in Fig.~\ref{Interactions}, gives four low-energy modes at the momenta $k=0,\pm k_F$,  where $k_F =2m \alpha_R$. Correspondingly, for small energies, we can split the field operators up into four modes,
\begin{eqnarray}
\psi_\uparrow (x) &\approx e^{ik_F x}  \psi_{R \uparrow}(x) + \psi_{L \uparrow}(x), \\
\psi_\downarrow (x) &\approx \psi_{R \downarrow}(x) + e^{-ik_F x} \psi_{L \downarrow}(x).
\end{eqnarray}
Next, we express $\psi_{\alpha s}(x)$ for $\alpha \in \{L,R\}$ and $s \in \{ \uparrow,\downarrow \}$ in terms of its Fourier components,
\begin{equation}
\psi_{\alpha s}(x) = \frac{1}{\sqrt{L}} \sum_k e^{i k x} \psi_{\alpha s,k},
\end{equation}
where $L$ is the length of the wire. In terms of these operators, the interaction Hamiltonians after projection to the lowest subband can be written as follows. The density-density interaction term reads
\begin{equation}
V_\rho = \frac{\tilde{V}(0)}{L} {\sum_{\alpha,s}} \int \, dx \, \rho_{\alpha s} (x) \rho_{\alpha s} (x),
\label{densityterms}
\end{equation}
where the fermionic densities are defined as usual as $\rho_{\alpha s}(x) = \psi^\dagger_{\alpha s}(x) \psi_{\alpha s}(x)$ and $\tilde{V}(q)$ is the Fourier transform of the interaction potential $V(\vec{r})$ projected to the lowest subband.

Due to momentum conservation, the spin-flip terms only mix terms near $k=0$,
\begin{equation}
V_{\rm{sf}} = v_{\rm{sf}} \int \, dx \left[ \psi^\dagger_{R\uparrow} (\partial_x \psi^\dagger_{R\uparrow}) (\partial_x \psi_{L\downarrow}) \psi_{L\downarrow} + \rm{h.c.} \right],
\label{SpinFlip}
\end{equation}
where we retained only the leading local part $v_{\rm{sf}} = -2\tilde{U}(0) \propto m \alpha_R^4 /\omega^3$, and $\tilde{U}(q)$ is given in Eq.~(\ref{uofq}). The full calculational details can be found in Appendix A. It is the Hamiltonian $V_{\rm{sf}}$ which allows for spin-umklapp scattering.

Most of the spin-exchange terms in $V_{\rm{sx}}$ can be expressed as density-density interactions, leading merely to changes in the coefficients of the terms in Eq.~(\ref{densityterms}). We separate out the single non-density-density term
\begin{equation}
V_{\rm{S}} = v_{\rm{S}} \int \, dx \biggl( \psi_{L\downarrow}^\dagger \psi_{R\uparrow}^\dagger \psi_{L\uparrow} \psi_{R\downarrow} + \rm{h.c.}  \biggr),
\label{SDWTerm}
\end{equation}
which corresponds to an interaction between inner and outer bands with strength $v_{\rm{S}} = 2k_F^2 \tilde{U}(k_F)$. In the limit in which this term dominates, it results in a spin-density wave state at $q=2k_F$, and has been discussed in detail in Ref.~\cite{Starykh+2008}.

Finally, we allow for the possibility of proximity-induced coupling to an $s$-wave superconductor, which pairs spin-up and spin-down electrons, and so has an effect at all the Fermi points in Fig.\ref{Interactions}. This pairing contribution to the Hamiltonian is
\begin{equation}
V_{\rm{SC}}= v_{\rm{SC}} \int \, dx \left( \psi_{R\uparrow}^\dagger \psi_{L\downarrow}^\dagger + \psi_{L\uparrow}^\dagger \psi_{R\downarrow}^\dagger + \rm{h.c.} \right),
\end{equation}
where $v_{\rm{SC}}$ is determined by the strength of the proximity-coupling to the superconductor. We now analyse the competition between these three possible interaction channels, parametrised by $v_{\rm{sf}}$, $v_{\rm{S}}$ and $v_{\rm{SC}}$.

\section{Bosonization \& renormalization group (RG) analysis.}
To further analyze the interacting system, we write the Hamiltonian in terms of bosonic operators $\phi_{\pm}$ and $\theta_{\pm}$ by defining
\begin{eqnarray}
\psi_{R \uparrow} &=  \frac{\eta_{R \uparrow} }{\sqrt{2\pi a}} e^{-i(\phi_+ - \theta_+)}, \hspace{20mm} \psi_{R \downarrow} =  \frac{ \eta_{R \downarrow}}{\sqrt{2\pi a}} e^{-i(\phi_- - \theta_-)},\nonumber \\
\psi_{L \uparrow} &= \frac{\eta_{L \uparrow} }{\sqrt{2\pi a}}  e^{i(\phi_- + \theta_-)}, \hspace{20mm} \psi_{L \downarrow} =  \frac{ \eta_{L \downarrow} }{ \sqrt{2\pi a}} e^{i(\phi_+ + \theta_+)},
\end{eqnarray}
where $\eta_{\alpha \sigma}$ are Klein factors and $a$ denotes the short-distance cutoff. Here, $\phi_+(x)$ and $\theta_+(x)$ are canonically conjugate bosonic operators for degrees of freedom near $k = \pm k_F$, whereas $\phi_-(x)$ and $\theta_-(x)$ describe modes near $k=0$. In these variables, the Hamiltonian consists of two Luttinger Hamiltonians for the $+$ and $-$ species, with approximately equal Luttinger parameters $K_\pm$, and interaction terms reflecting Eqs.~(\ref{SpinFlip}) and (\ref{SDWTerm}). Moreover, one obtains derivative terms which couple the two species $\kappa_\phi \partial_x \phi_+ \partial_x \phi_-$ and $\kappa_\theta \partial_x \theta_+ \partial_x \theta_-$.

Following Ref.~\cite{Starykh+2008}, we can diagonalize the quadratic parts of the Hamiltonian by going to the charge-spin basis $\phi_{\rho,\sigma} = (\phi_+ \pm \phi_-)/\sqrt{2}$ and $\theta_{\rho,\sigma} = (\theta_+ \pm \theta_- )/\sqrt{2}$ so that
\begin{equation}
H_0 = \sum_{a=\rho,\sigma} \frac{v_a}{2 \pi} \int \, dx \left[ \frac{(\partial_x \phi_a)^2}{K_a} + K_a (\partial_x \theta_a)^2 \right],
\label{NonInteractingHamiltonian}
\end{equation}
where $v_{\rho,\sigma}$ are the respective sound velocities of the modes. For repulsive interactions, we have $K_\rho < 1$ and $K_\sigma < 1$ \cite{Starykh+2008}. In addition to $H_0$, we obtain two competing interaction terms,
\begin{eqnarray}
    V_{\rm{S}} &= \frac{g_{\rm{S}}}{(2 \pi a)^2} \int \, dx \, \cos [2\sqrt{2} \theta_\sigma], \\
    V_{\rm{sf}} &= \frac{ g_{\rm{U}}}{(2 \pi a)^2} \int \, dx \, \cos [ 2\sqrt{2}(\phi_\rho-\phi_\sigma)].
\end{eqnarray}
Proximity-induced $s$-wave superconductivity gives a contribution which reads in bosonized form,
\begin{equation}
V_{\rm{SC}} = \frac{ g_{\rm{SC}}}{(2 \pi a)^2} \int \,dx \, \biggl( \cos[\sqrt{2}(\theta_\rho + \theta_\sigma)] + \{\theta_\sigma \rightarrow -\theta_\sigma\} \biggr).
\end{equation}
For weak interactions, all parameters of the model can be determined precisely in the bosonization procedure (see Appendix A). However, for strong interactions it is more convenient to regard the parameters $v_{\rho,\sigma}$ and $K_{\rho,\sigma}$ as well as the three coupling strengths $g_{\rm{S}}$, $g_{\rm{U}}$ and $g_{\rm{SC}}$ as effective parameters, which may flow independently under renormalization as we change the cut-off $a$.

We calculate the flow of the various coupling constants using real-space RG calculation based on operator product expansions \cite{cardy96}. We find the following first-order RG equations for the coupling constants of the cosine terms (see Appendix B for details of the RG proceedure),
\begin{eqnarray}
    \frac{d g_{\rm{S}}}{d \ell} &= \left( 2-\frac{2}{K_\sigma} \right) g_{\rm{S}}, \label{RGsdw}  \\
    \frac{d g_{\rm{U}}}{d \ell} &=  2(1-K_\sigma-K_\rho) g_{\rm{U}}, \label{RGumklapp}     \\
    \frac{d g_{\rm{SC}}}{d\ell} &= \left( 2-\frac{1}{2K_\sigma} - \frac{1}{2K_\rho} \right) g_{\rm{SC}}, \label{RGsuperconductivity}
\end{eqnarray}
implying that the spin-density wave term is always irrelevant for repulsive interactions ($K_\sigma < 1$) \cite{Starykh+2008}. The spin-umklapp term, by contrast, can become relevant for strong interactions where $K_\rho + K_\sigma < 1$. Finally, the superconducting term is relevant for $K_\rho^{-1} + K_\sigma^{-1} < 4$.

We would like to point out that for $\alpha_R =0$, the system becomes $\rm{SU}(2)$ invariant. In that case, the spin-umklapp term vanishes because spin is conserved and one finds the well-known Kosterlitz-Thouless RG flow which brings $K_\sigma \to 1$ as $g_{\rm{S}} \to 0$. In contrast, for $\alpha_R \neq 0$, $K_\sigma$ is not constrained and strong repulsive interactions lead to $K_\sigma \ll 1$.
	
We start the RG flow from an initial value $a = a_0$ and flow towards $a \sim L$, the length of the wire. Generically, the RG flow will stop at a finite value $a_\infty < L$ as soon as one of the dimensionless coupling constants $g_{\rm{SC,U}}$ approaches one. The bare value of $g_{\rm{SC}}(a_0)$ is determined by the strength of the proximity coupling to the superconductor, which can be experimentally optimized. The bare value $g_{\rm{U}}(a_0)$ depends on the separation between the lowest subbands, and so depends on the transverse confinement (i.e.~the physical width) of the wire.

To generate zero-energy bound states, spin-umklapp scattering must gap out the modes near $k=0$, whereas proximity-induced superconductivity should open a gap for the modes at $k=\pm k_F$, see Fig.~\ref{Interactions}. Superconductivity affects all modes, so this is only possible if at the end of the RG flow $|g_{\rm{U}}(a_\infty)| > |g_{\rm{SC}}(a_\infty)| > 0$. Strong electron-electron interactions result in $K_\rho <1/2$ and $K_\sigma<1/2$, which a priori makes the spin-umklapp term relevant and the superconducting term irrelevant. However, since the RG flow is cut off at a finite length scale, one will generally find a nonzero $|g_{\rm{SC}}(a_\infty)| > 0$ at the end of the RG flow, meaning that a superconducting gap will still open.
 		
To demonstrate the existence of degenerate, localized zero-energy bound states, we use the unfolding transformation described in Refs.~\cite{Oreg+2014,Giamarchi+2004}. This transformation can be used to map our system with length $L$ and open boundary conditions to a system of length $2L$ and periodic boundary conditions. Explicitly, we construct the unfolded \emph {chiral} fields
\begin{eqnarray}
\xi_R(\tx) = \left\{\begin{array}{ll}
\varphi_{R+} (\tx) & {\rm{for}} \  0 \leq \tx \leq L\\
\varphi_{L-} (2L-\tx) & {\rm{for}} \,\, L \leq \tx \leq 2L
\end{array}\right.\\
\xi_L (\tx) = \left\{\begin{array}{ll}
\varphi_{L+} (\tx) & {\rm{for}} \ 0 \leq \tx \leq L \\
\varphi_{R-} (2L-\tx) & {\rm{for}} \ L \leq \tx \leq 2L
\end{array}\right.
\end{eqnarray}
where $\varphi_{\alpha \nu} = \alpha \phi_{\nu} - \theta_{\nu}$ with $\alpha=R,L$ and $\nu = \pm$. In order that the fermionic fields obey the vanishing boundary conditions at $x=0$ and $x=L$, we find that the bosonic fields must obey
\begin{eqnarray}
\varphi_{L+} (0) = \varphi_{R-}(0), & \qquad \varphi_{L+} (L) = \varphi_{R-}(L)\nonumber \\
\varphi_{L-} (0) = \varphi_{R+}(0), & \qquad \varphi_{L-} (L) = \varphi_{R+}(L)
\end{eqnarray}
Note that in this transformation, the degrees of freedom $(\phi_+,\theta_+)$ are mapped on the range $\tx \in [0,L]$, whereas the $(\phi_-,\theta_-)$ fields are mapped on the range $\tx \in [L,2L]$. Since the original chiral fields satisfy $[\varphi_{\alpha\nu}(x),\varphi_{\alpha^\prime\nu^\prime}(x')] =  i \pi \alpha\delta_{\alpha \alpha^\prime} \delta_{\nu \nu^\prime} {\rm{sgn}}(x-x')$, we find that the unfolded fields obey the correct chiral commutation relations
\begin{equation}
[\xi_{\alpha}(\tx),\xi_{\alpha'}(\tx')] = i\pi \alpha \delta_{\alpha\alpha'}{\rm{sgn}}(\tx-\tx'),
\end{equation}
on the whole interval $\tx,\tx' \in [0,2L]$. In terms of these unfolded fields, our Hamiltonian for the relevant perturbations arising from umklapp scattering and superconductivity reads
\begin{eqnarray}
V_{\rm{sf}} + V_{\rm{SC}} &=& \int_0^{2L} d\tx \left[ g_{\rm{U}}(\tx) \cos(2[\xi_R(\tx)-\xi_L(\tx)]) + g_{\rm{SC}}(\tx) \cos(\xi_R(\tx)+\xi_L(\tx)) \right] \nonumber \\
& &
\end{eqnarray}
where the position-dependent couplings $g_{\rm{U}}(\tx)$ and $g_{\rm{SC}}(\tx)$ have support on $\tx\in [0,L]$ and $\tx\in [L,2L]$ respectively. The unfolded system consists of two adjacent regions. Between $\tx=0$ and $\tx=L$, we have a region of superconductor where the field $\theta(\tx)=-(\xi_R+\xi_L)/2$ is pinned by the term $\cos[2\theta]$ to the value $\theta^\star = (n+1/2)\pi$ for integer $n$. From $\tx=L$ to $\tx=2L$, there is a region of Mott insulator where the field $\phi=(\xi_R-\xi_L)/2$ is pinned by the term $\cos[4 \phi]$ to the value $\phi^\star=1/2(m+1/2)\pi$ for integer $m$. The spectrum is completely gapped, except possibly at the boundaries between the two regions, $\tx = 0$ and $\tx = L$, where the parafermion states we describe emerge. The unfolded system is identical to the topological insulator edge state system studied in Ref.~\cite{Orth+2015}, which tells us that our original system contains $\mathbb{Z}_4$ parafermion state at its ends.

Let us reproduce the essential parts of the derivation here. We define the total charge and total spin operators for the system, according to
\begin{eqnarray}
\pi S &= \theta(2L)-\theta(0), \\
\pi Q &= \phi(2L)-\phi(0).
\end{eqnarray}
Despite the fact that the fermionic fields must be continuous, the bosonic fields $\phi$ and $\theta$ may jump by integer multiples of $2\pi$, so that $S$ and $Q$ can be nonzero in spite of the periodic boundary conditions. The spin $S$ of the system takes integer values, and is conserved $mod(4)$ so  $s=\{0,1,2,3\}$ (measured in units of $\hbar/2$), whereas the charge $Q$ takes half-integer values and is conserved $mod(2)$ so $q=\{0,\frac{1}{2},1,\frac{3}{2} \}$ (measured in units of $e$). Since we may only add integer amounts of electronic charge to our junction, we must restrict our value of the charge to be $q\in \{0,1\}$ \footnote{Note that in a system where there are several junctions between superconducting and Mott insulating regions, it is perfectly acceptable to have states with half-integer charge, as long as the total charge of the complete system is restricted again to $q\in \{0,1\}$.}. The state of our system is then defined by $|s,q \rangle$. Since every physical electron carries one unit of spin, this means that for the charge state $q=0$, only the two total spin states $s\in \{0,2\}$ are permissible. Similarly, $q = 1$ requires $s \in \{1,3\}$. Hence, we have a total fourfold degeneracy of the ground state.

To see explicitly the parafermionic statistics of the bound states at $x=0$ and $x=L$, we write the pinned values of the fields in terms of integer-spectrum operators $m,n_1,n_2$ as
\begin{eqnarray}
\phi_1 &=& \frac{1}{2} \left( \pi m + \frac{\pi}{2} \right), \nonumber \\
\theta_{1,2} &=&  \pi n_{1,2} + \frac{\pi}{2}.
\end{eqnarray}
These operators then have commutation relations
\begin{equation}
[m,n_1] = -\frac{i}{\pi},
\end{equation}
\begin{equation}
[m,n_2] = 0.
\end{equation}
The total spin operator is given by
\begin{equation}
\hat{S} = e^{i \pi m/2} = e^{-i\pi/4} e^{i \phi_1}
\end{equation}
and the total spin in the system is $\langle \hat{S} \rangle = (\theta_1-\theta_2)/\pi = n_2-n_1$. Then the parafermion states obeying
\begin{equation}
\chi_1 \chi_2 = e^{-i\pi/2} \chi_2 \chi_1
\end{equation}
are given by
\begin{eqnarray}
\chi_1 = T_Q e^{i \pi m/2}, \nonumber \\
\chi_2 = e^{i\pi/4} T_Q e^{i \pi m/2} e^{i \pi (n_2-n_1)/2},
\end{eqnarray}
where $T_Q$ is the raising operator for charge $T_Q |s,q \rangle = |s,q+1 \rangle$.
	
\section{Josephson effect \& Shapiro steps.}
A zero-bias conductance peak is a possible experimental signature of localized Majorana fermions. However, such peaks could arise from other mechanisms, e.g., disorder \cite{Pikulin+2012,Liu+2012}, and do not directly indicate the degeneracy of the states involved. In particular, the fourfold degenerate bound state we propose would lead to the same zero-bias anomaly in transport measurements, albeit at vanishing magnetic field. To uniquely discriminate these particular bound states, we instead propose to discover their presence via the periodicity of the Josephson effect, similar to the corresponding proposal for Majorana fermions \cite{Fu+2009,rokhinson12}.

\begin{figure}[t]
\centering
\includegraphics[width=0.45\linewidth]{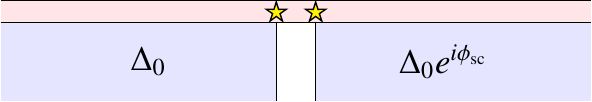}
\caption{Experimental setup for the measurement of the $8 \pi$ periodic Josephson effect. Two superconductors underneath a Rashba wire are held at a phase difference $\phi_{\rm{sc}}$. Fractionally charged bound states are indicated by stars.}
	\label{SCSetup}
\end{figure}

To investigate the effect of the zero-energy bound states on the Josephson effect, we follow the logic of Ref.~\cite{Zhang+2014,Orth+2015} and consider an arrangement with two superconducting contacts with phase difference $\phi_{\rm{sc}}$ placed under a Rashba wire partially gapped by spin-umklapp scattering (see Fig.~\ref{SCSetup}), in an analogous arrangement to the experimental setup of Ref.~\cite{rokhinson12}. The wire adjacent to the edges of the superconductors will host zero-energy modes with charge $e/2$, which will dominate the transport at low energies and for a short junction. Tunnelling of a single quasiparticle through the junction changes the parity of the end states. In order to satisfy the boundary conditions due to the applied superconducting phase, \emph{four} $e/2$ quasiparticles must tunnel via the bound states, leading to an $8\pi$ periodic Josephson effect \cite{Zhang+2014}. The time-reversal breaking of the superconducting phase difference causes a slight lifting of the fourfold degeneracy, but for realistic parameters this shift is negligible \cite{Zhang+2014}.

Several works \cite{Fidkowski+2011,turner11} have suggested that in a spinless, one-dimensional system, the greatest achievable topological degeneracy is twofold, leading to the statement that only Majorana fermions can exist in one-dimensional systems. Our system does not contradict this theorem because in our case the ground state degeneracy is not entirely topological. Indeed, it can be viewed as a twofold topological degeneracy combined with a twofold degeneracy due to time-reversal symmetry \cite{Sela+2011}. This second degeneracy can be lifted by local TR symmetry-breaking perturbations such as a magnetic field. In that case, only the topological part of the ground state degeneracy survives, and one recovers the $4\pi$ periodicity of the Josephson effect seen for Majorana bound states \cite{Fu+2009} \footnote{The two possible states of the parafermion at each end of the wire are Kramers partners, and so are automatically protected from any time-reversal invariant local perturbations at the edges, so we need only concern ourselves with time-reversal symmetry breaking disorder, and non-magnetic bulk disorder.}. In real material samples, time reversal symmetry may also be weakly broken by magnetic impurities, thereby lifting the fourfold degeneracy to a twofold one, even for a very long wire. In either case, the undriven Josephson current is no longer $8\pi$ periodic. This raises the question of whether remnants of the $8 \pi$ periodicity can be observed in such a nonideal setting.

A possible answer was proposed for similar problems in Majorana nanowires \cite{Dominguez+2012}, where a $4 \pi$ periodicity is reduced by parity-flipping perturbations to a trivial $2\pi$ periodicity: by driving the current in the junction at a finite frequency, we allow Landau-Zener tunneling between the different low-lying states.  Then, Shapiro step measurements \cite{Shapiro+1963} can still distinguish higher periodic components even when those are very weak, as in the case of a $4\pi$ periodicity recently reported in experiments on Majorana bound states \cite{rokhinson12, Wiedenmann+2015}.

In our finite-length Rashba wire, there will generically be a nonzero overlap of the modes at each end of the wire and so the degeneracy between the modes will be split, although this effect is exponentially suppressed in the length of the wire. However, in the case of a driven junction, Landau-Zener tunneling gives us access to all the low energy modes, even when they are subject to a small splitting. Allowing for the possibility of Josephson tunneling of Cooper pairs, and for the tunneling of Majorana fermions and $\mathbb{Z}_4$ parafermions through the weak link in our system, the total Josephson current flowing is given by
\begin{equation}
I(\varphi(t)) = i_c \sin (\varphi(t)) +i_m \sin (\varphi(t)/2) + i_p \sin (\varphi(t)/4).
\end{equation}
The amplitude $i_c$ accounts for the current due to the tunneling of Cooper pairs (this is the critical current above which the junction becomes ``normal''). The parameters $i_m$ and $i_p$ similarly account for the tunneling of Majorana fermions and $\mathbb{Z}_4$ parafermions through the junction. The Josephson equation relating the rate of change of the superconducting phase $\varphi$ to the voltage across the junction reads
\begin{equation}
\dot{\varphi}(t) = \frac{2e}{\hbar} V(t).
\end{equation}
We \emph{current-bias} our Josephson junction using a constant current with a small a.c.~component $I_0 + I_1 \sin (\omega t)$. The current through the junction consists of two parallel components, a \emph{tunnel current} given by $I(\varphi(t))$ and a \emph{resistive current} due to ohmic quasiparticle transport in the junction. Equating the sum of these two contributions to the bias current, the gauge-invariant phase $\varphi(t)$ is described by the dynamical equation
\begin{equation}
I_0 + I_1 \sin (\omega t) = i_c \sin (\varphi(t)) +i_m \sin (\varphi(t)/2) + i_p \sin (\varphi(t)/4) + \frac{\hbar}{2eR} \dot{\varphi}(t).
\label{Jcurrent}
\end{equation}
Note that these contributions to the Josephson current are periodic under shifts $\varphi(t) \rightarrow \varphi(t)+2 \pi$, $\varphi(t) \rightarrow \varphi(t)+4 \pi$ and $\varphi(t) \rightarrow \varphi(t)+8 \pi$ respectively. Writing the equation (\ref{Jcurrent}) in terms of rescaled variables $\tau=2eR i_c t/\hbar$, and $\tilde{\omega} = \hbar \omega/2eR i_c$, we find the equation
\begin{equation}
 \dot{\varphi}(\tau) = \alpha_0 + \alpha_1 \sin (\tilde{\omega} \tau) -\sin (\varphi(\tau)) -\alpha_m \sin (\varphi(\tau)/2)- \alpha_p \sin (\varphi(\tau)/4)
\label{Jcurrent2}
\end{equation}
This equation must be solved numerically. 

\begin{figure}[t]
	\centering
	\includegraphics[width=0.95\linewidth]{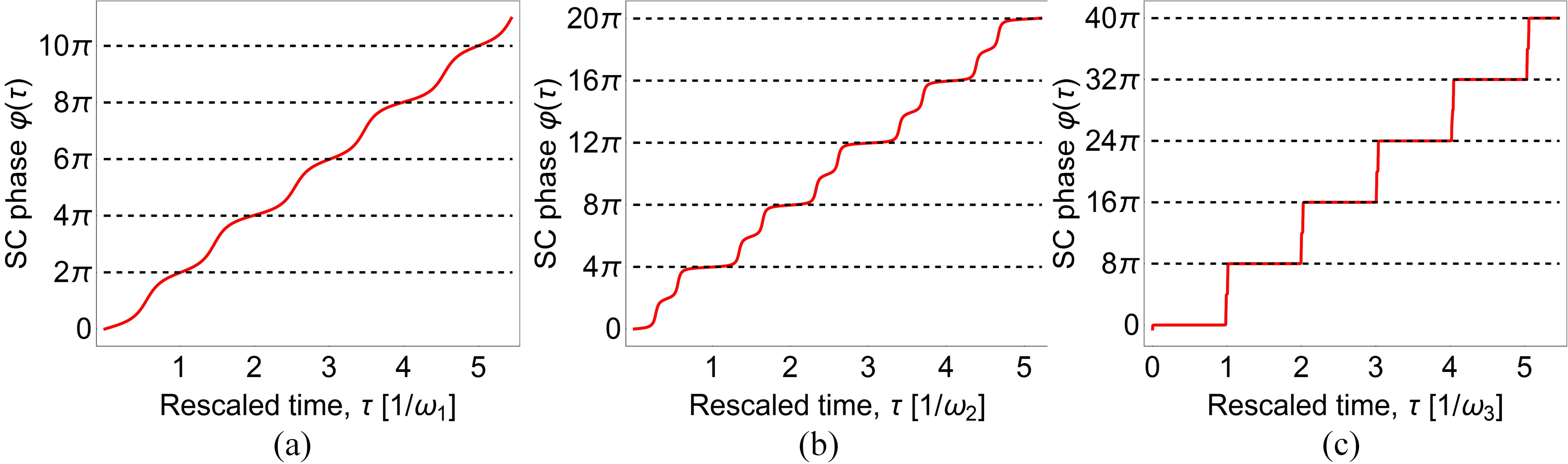}
	\caption{Numerical solutions to Eq.(\ref{Jcurrent2}) with a d.c. bias current, showing different periodicity behaviour depending on the amplitude of the d.c. current. We have made the generic choice of $i_c=10i_m=100i_p=1$ for the relative size of the contributions to the Josephson current from Cooper pairs, Majorana fermions and parafermions. Panel (a) shows the superconducting phase changing in $2\pi$ steps for the choice $\alpha_0=1.5$, whereas panel (b) is plotted for $\alpha_0=1.1$, and shows residual small steps at $2\pi$ and dominant steps at $4\pi$ which corresponds to the dominant transport being via tunneling of Majorana fermions. In panel (c), we choose $\alpha_0=1.08$, and recover jumps of $8\pi$ in the superconducting phase due to tunneling of parafermions.}
	\label{FigNoDriveCurrent}
\end{figure}

In order to see that the $8\pi$ periodicity can win out, even when $\alpha_c=1>\alpha_m>\alpha_p$, we first solve (\ref{Jcurrent2}) without the a.c. drive current, i.e. $\alpha_1=0$ (see Fig.~\ref{FigNoDriveCurrent}). For a generic choice of the d.c. bias current, $\alpha_0$, the superconducting phase is $2\pi$ periodic, reflecting the dominance of the Cooper pair tunneling (Fig.~\ref{FigNoDriveCurrent}a). However, we find that as we approach a critical value of $\alpha_0$, we first see the $4\pi$ periodic term due to the Majorana contribution (Fig.~\ref{FigNoDriveCurrent}b) and then the $8\pi$ contribution from the parafermions (Fig.~\ref{FigNoDriveCurrent}c) become dominant.

\begin{figure}[t]
	\centering
	\includegraphics[width=0.65\linewidth]{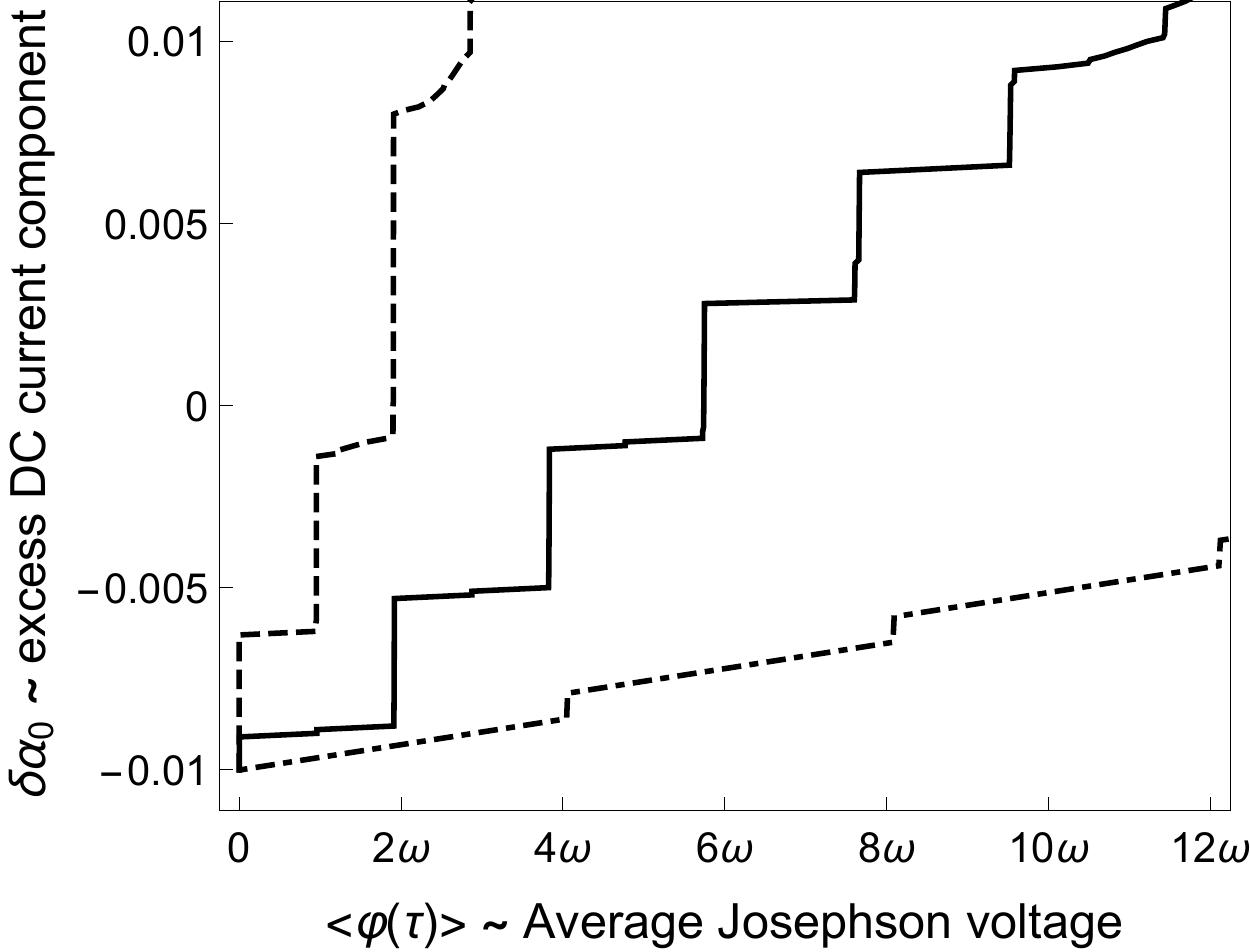}
	\caption{A plot of the current-voltage relationship for the system driven with a small amplitude a.c. current with a large d.c. offset (corresponding to $\alpha_0=1.08+\delta \alpha_0$ and $\alpha_1=0.01$). The excess current $\delta \alpha_0$ over the critical value at $\tilde{w}=0$ is plotted on the vertical axis. The Josephson junction has a dominant $2\pi$ periodic component, and sub-leading $4\pi$ and $8\pi$ periodic components ($i_c=10i_m=100i_p=1$). The dashed curve shows the system driven at a frequency $\tilde{\omega}_1=0.04$ (comparable to that of the $2\pi$ periodic component of the Josephson current), and shows all integer Shapiro steps. The solid line is the system driven with an a.c. component with frequency $\tilde{\omega}_2=0.01$, and only even Shapiro steps survive. Finally, the dashed-dotted line shows the $\tilde{\omega}_3=0.00052$ behaviour, and the disappearance of all the Shapiro steps which are not a multiple of 4. Note that on the same scale as the $\tilde{\omega}_1$ and $\tilde{\omega}_2$ plots, the $\tilde{\omega}_3$ shows very small variation, so its amplitude has been increased by a factor of $7$ to clearly show the disappearance of the Shapiro steps.}
	\label{shapirostepsfig}
\end{figure}

The winding up of the superconducting phase is not a property which is easy to directly measure, so to give an experimentally accessible measurement, we must drive the Landau-Zener transitions which will allow us to see the degeneracy of the low-lying modes. To do this, we switch on the small a.c. component to the bias current $\alpha_1$ (experimentally achieved by irradtiating the junction with microwaves). Tuning the driving frequency allows us to access three distinct regimes, in which steps occur in the experimentally-measurable I-V curves for the junction at distinct multiples of the driving frequency $\tilde{\omega}$. For high frequencies, we find Shapiro steps are present at \emph{all} integer multiples of the driving frequency, indicating that the transport is dominated by conventional Josephson tunneling of Cooper pairs through the junction. Reducing the frequency of the drive, we recover the results of Ref.~\cite{Dominguez+2012}, that only the \emph{odd}-numbered Shapiro steps remain. Finally, for very low frequencies, there exists a regime in which only every $(4n+1)^{th}$ Shapiro step survives (see Fig. \ref{shapirostepsfig}). Whilst the appearance of extra Shapiro steps can be caused e.g.~by disorder, alternative mechanisms to the one described by which Shapiro steps may \emph{disappear} seem to be unknown. As the disappearing Shapiro steps are robust even when the $2\pi$ and $4\pi$ contributions to the Josephson current are dominant over the $8 \pi$ component, they therefore provide a highly selective test of the existence of $\mathbb{Z}_4$ parafermions.

\section{Discussion.} 

	
To summarize, we predict the existence of fourfold degenerate zero-energy bound states protected by time-reversal symmetry when inducing superconductivity in strongly interacting Rashba wires. The bound state operators acting in the ground state manifold satisfy a $\mathbb{Z}_4$ parafermionic algebra, and transitions between the degenerate ground states occur via the tunneling of fractional charges $e/2$. We propose an experimental scheme allowing us to observe an $8\pi$ periodic Josephson current and associated Shapiro step structure, which would be a definitive signature of our predicted bound states.

The opening of the spin-umklapp gap at $k=0$ is rather generic. In a wire with RSOC and electron-electron interactions this gap will lead to a reduction of the normal-state conductance of from $2e^2/h$ to $e^2/h$ as the chemical potential approaches the Dirac point. A similar phenomenon has been observed in GaAs wires \cite{Zumbuhl+2014} and our results may provide an alternative interpretation for this experiment if RSOC is sufficiently strong in the semiconductor nanowires used. A conductance measurement as a function of the chemical potential in InSb or InAs wires, where RSOC is typically stronger, could demonstrate the proposed reduction of the conductance conclusively. We would like to point out that the required strong interactions have already been seen in these wires \cite{Hevroni+2015}. Umklapp scattering has also recently been invoked to explain the observed conductance reduction in InAs/GaSb topological insulator edge states \cite{li15}. Introducing a weak superconducting proximity effect in either of these types of wires or edge states will then lead to the creation of bound states and allow the observation of the $8\pi$ periodic Josephson current component.

\section{Acknowledgements.}
 We would like to thank Peter Schmitteckert and Alexander Zyuzin for helpful discussions. TLS \& CJP are supported by the National Research Fund, Luxembourg under grant ATTRACT 7556175. TM is funded by Deutsche Forschungsgemeinschaft through GRK 1621 and SFB 1143. RPT acknowledges financial support from the Swiss National Science Foundation.

\appendix
\section{Interactions}

The interaction Hamiltonian for a density-density interaction reads
\begin{equation}
	\hat{V} = \sum_{\sigma_1 \sigma_2} \int \, d^2r_1 d^2r_2   \psi_{\sigma_1}^\dagger (\vec{r}_1) \psi_{\sigma_2}^\dagger (\vec{r}_2) V (|\vec{r}_1-\vec{r}_2|) \psi_{\sigma_2}(\vec{r}_2) \psi_{\sigma_1}(\vec{r}_1).
	\label{PotentialTerms}
\end{equation}	
and the operator $\psi_\sigma^\dagger(\vec{r})$ creates a particle with spin $\sigma \in \{ \uparrow,\downarrow\}$ at position $\vec{r} = (x,y)$. We transform this term using the Schrieffer-Wolff transformation, and then project onto the lowest subband. Due to translation invariance, the momentum in $x$ direction remains a good quantum number. The Hamiltonian in momentum space in this sector is
\begin{equation}
	\hat{H} = \hat{H}_{0} + \hat{V}_\rho + \hat{V}_{\rm{sf}} + \hat{V}_{\rm{sx}},
\end{equation}
where the kinetic energy takes the form
\begin{equation}
	\hat{H}_{0} = \sum_{\sigma,p} \left( \frac{p^2}{2m} + \alpha_{\rm{R}} \sigma p \right) \psi^\dagger_{p,\sigma} \psi_{p,\sigma},
\end{equation}
and the operators $\psi^\dagger_{p,\sigma}$ and $\psi_{p,\sigma}$ create and annihilate an electron of spin $\sigma$ and momentum $p$. The potential terms are then $\hat{V}_\rho$, and the density-density interaction becomes
\begin{equation}
	\hat{V}_\rho = \frac{1}{L} \sum_{\sigma_1 \sigma_2} \sum_{p,p^\prime,q} \tilde{V} (q) \psi^\dagger_{p+q,\sigma_1} \psi^\dagger_{p^\prime-q,\sigma_2} \psi_{p^\prime,\sigma_2} \psi_{p,\sigma_1},
\end{equation}
where we have introduced $\tilde{V}(q,y)$ as the partial Fourier transform in $x$ direction of the physical interaction potential $V(|\vec{r}|) \equiv V(x,y)$,
\begin{equation}
	\tilde{V}(q,y) = \int dx e^{-i q x} V(x, y)
\end{equation}
and defined $\tilde{V}(q) \equiv \tilde{V}(q,y=0)$. Next, the spin flip and spin exchange terms are given by
\begin{eqnarray}
	\hat{V}_{\rm{sf}} &= \frac{1}{L} \sum_{\sigma} \sum_{p,p^\prime,q} \tilde{U} (q) (2p^\prime - q)(2p+q) \psi^\dagger_{p+q,\sigma} \psi^\dagger_{p^\prime-q,\sigma} \psi_{p^\prime,-\sigma} \psi_{p,-\sigma}, \\
	\hat{V}_{\rm{sx}} &= \frac{1}{L} \sum_{\sigma} \sum_{p,p^\prime,q} \tilde{U} (q) (2p^\prime - q)(2p+q) \psi^\dagger_{p+q,\sigma} \psi^\dagger_{p^\prime-q,-\sigma} \psi_{p^\prime,\sigma} \psi_{p,-\sigma},
\end{eqnarray}
with an effective potential $\tilde{U}(q)$ given by
\begin{equation}
	\tilde{U}(q) = \frac{m \alpha_R^4}{2 \pi^{3/2}\omega^3}
	\sum_{n=0}^\infty \frac{1}{\sqrt{2^n n!}} \int_{-\infty}^\infty dz_1 dz_2 dz_3 e^{-z_1^2-z_2^2-\frac{(z_1-z_2)^2}{2}}  H_{n}(z_1 - z_2) H_1(z_1) H_1(z_2) e^{-z_3^2/2} H_n(z_3) \tilde{V} \left(q, \frac{z_3}{\sqrt{m \omega}} \right),
	\label{uofq}
\end{equation}
where $H_n(z)$ denote Hermite polynomials which emerge because they are the transversal eigenfunctions due to the harmonic confinement in $y$ direction.

We now project these terms onto the linearized low-energy states of the system. For the density-density term, we obtain
\begin{equation}
	\hat{V}_\rho = \tilde{V}(0) \sum_{\alpha,\sigma} \int \, dx \, \rho_{\alpha,\sigma} (x) \rho_{\alpha,\sigma} (x)
\end{equation}
with the fermionic densities $\rho_{\alpha \sigma}(x) = \psi^\dagger_{\alpha \sigma}(x) \psi_{\alpha \sigma}(x)$. Due to momentum conservation in $x$ direction, the spin flip terms may only mix terms near $q=0$, so $\hat{V}_{\rm{sf}}$ reads
\begin{equation}
	\hat{V}_{\rm{sf}} = -2\tilde{U} (0) \int \, dx \left( \psi^\dagger_{R\uparrow} (\partial_x \psi^\dagger_{R\uparrow}) (\partial_x \psi_{L\downarrow}) \psi_{L\downarrow} + \rm{h.c.} \right)
\end{equation}
Finally, the spin exchange terms only lead to a small change in terms already present in $\hat{V}_\rho$, so we can account for them by a change of the interaction parameters, but we should keep separate the one term which may not be written as a density-density interaction;
\begin{equation}
	\hat{V}_{\rm{S}} = 2k_F^2 \tilde{U} (k_F) \int \, dx \biggl( \psi_{L\downarrow}^\dagger \psi_{R\uparrow}^\dagger \psi_{L\uparrow} \psi_{R\downarrow} \nonumber + \psi_{R\downarrow}^\dagger \psi_{L\uparrow}^\dagger \psi_{R\uparrow} \psi_{L\downarrow}  \biggr).
\end{equation}
We now bosonize these terms. Using the standard identity that $\rho_{\alpha,\sigma} = \partial_x \phi_{\alpha,\sigma}/2\pi$, we find that at weak coupling, the complete interaction Hamiltonian $\hat{V} = \hat{V}_\rho+\hat{V}_{\rm{sf}}+\hat{V}_{\rm{sx}}$ becomes
\begin{eqnarray}
	\hat{V} &=\frac{ 4\tilde{V}(0)}{(2\pi)^2} \int \, dx   \left((\partial_x \phi_+)^2 + (\partial_x \phi_-)^2 \right) +\frac{1}{(2\pi)^2} \int \, dx \biggl(  \left[-8\tilde{V}(0)-2k_F^2\tilde{U}(k_F) \right] \partial_x \phi_+ \partial_x \phi_- + 2 k_F^2 \tilde{U}(k_F) \partial_x \theta_+ \partial_x \theta_- \biggr) \nonumber \\
	&- \frac{ 2 \tilde{U}(0)}{(2\pi a)^2} \int \, dx \, \biggl( e^{i(\phi_--\theta_-)} (\partial_x e^{i(\phi_--\theta_-)}) (\partial_x e^{i(\phi_-+\theta_-)}) e^{i(\phi_-+\theta_-)} + \rm{h.c.} \biggr) \frac{ 2k_F^2 \tilde{U}(k_F)}{(2\pi a)^2} \int \, dx \, \left( e^{i(2\theta_+ - 2\theta_-)} + \rm{h.c.} \right)
\end{eqnarray}

We note at this point that the Luttinger parameters for the two species are equal in the weak coupling limit. The coupling between the species given by the terms $\kappa_\phi \partial_x \phi_+ \partial_x \phi_-$ where $\kappa_\phi = -8\tilde{V}(0)-2 k_F^2\tilde{U}(k_F)$ and $\kappa_\theta \partial_x \theta_+ \partial_x \theta_-$ where $\kappa_\theta = 2 k_F^2 \tilde{U}(k_F)$. These can be removed by going to the charge-spin basis described in the text above Eqn.(10), which yields Eqn.(10) with new Luttinger parameters
\begin{eqnarray}
	K_\rho^2 &= \frac{K + \kappa_\theta/2}{1/K + \kappa_\phi/2}, \nonumber \\
	K_\sigma^2 &= \frac{K - \kappa_\theta/2}{1/K - \kappa_\phi/2},  
\end{eqnarray}
and renormalised Fermi velocities $v_\rho = v_F K_\rho$ and $v_\sigma = v_F K_\sigma$. $v_F = \alpha_{\rm{R}}$ is the Fermi velocity for the non-interacting modes. Note that these expressions are only valid at weak coupling. However, the division into a pair of non-interacting Luttinger Hamiltonians with three competing cosine interaction terms forms our prototypical model for the strongly interacting case we consider.

\section{Renormalization Group}

We calculate the scaling dimensions of the operators above, and also the second order RG flow for the parameters $K_\rho$ and $K_\sigma$. We diagonalize the non-interacting Hamiltonian (Eq.(10) in the main text) by introducing the fields $\varphi_{\alpha a} = \alpha \sqrt{K_a} \phi_a - \theta_a/\sqrt{K_a}$, where $a=\rho,\sigma$, and as before $\alpha=R,L$, which gives
\begin{equation}
	H_0 = \frac{v_\rho}{2\pi} \int \, dx \left( (\partial_x \varphi_{\rho R})^2 + (\partial_x \varphi_{\rho L})^2\right) + \frac{v_\sigma}{2\pi} \int \, dx \left( (\partial_x \varphi_{\sigma R})^2 + (\partial_x \varphi_{\sigma L})^2\right).
	\label{DiagonalHamiltonian}
\end{equation}
The scaling dimensions of the operators are calculated by normal-ordering them with respect to the creation and annihilation operators of the diagonal Hamiltonian (\ref{DiagonalHamiltonian}). We find
\begin{equation}
	V_{\rm{S}} = \frac{g_{\rm{S}}}{(2 \pi a)^2} \left( \frac{L}{2 \pi a} \right)^{-2/K_\sigma} \int \, :\cos [2\sqrt{2} \theta_\sigma]:.
\end{equation}
Asserting that this term cannot change as a result of the RG step $a\rightarrow a-da = a(1+d\ell)$ and $g_{\rm{S}} \rightarrow g_{\rm{S}} + dg_{\rm{S}}$ gives the first-order RG equation for $g_{\rm{S}}$ as
\begin{equation}
	\frac{d g_{\rm{S}}}{d \ell} = \left[ 2-\frac{2}{K_\sigma} \right] g_{\rm{S}}.
\end{equation}

Under normal ordering and point splitting, the umklapp term becomes
\begin{equation}
	V_{\rm{U}} = \frac{ g_{\rm{U}}}{(2 \pi a)^2} \left( \frac{L}{2 \pi a} \right)^{-2(K_\rho+K_\sigma)}  \int \, dx :\cos [ 2\sqrt{2}(\phi_\rho-\phi_\sigma)]:,
\end{equation}
and so has the first-order RG equation
\begin{equation}
	\frac{d g_{\rm{U}}}{d \ell} = \left[ 2(1-K_\sigma-K_\rho) \right] g_{\rm{U}}.
\end{equation}

The superconducting term has scaling behaviour
\begin{eqnarray}
	V_{\rm{SC}} &=& \frac{ g_{\rm{SC}}}{(2 \pi a)^2} \left( \frac{L}{2 \pi a} \right)^{-\frac{1}{2K_\rho} -\frac{1}{2K_\sigma}} \int \,dx \biggl( :\cos[\sqrt{2}(\theta_\rho + \theta_\sigma]:  + :\cos[\sqrt{2}(\theta_\rho - \theta_\sigma]: \biggr). \nonumber \\
	& &
\end{eqnarray}
giving an RG equation
\begin{equation}
	\frac{d g_{\rm{SC}}}{d\ell} = \left[ 2-\frac{1}{2K_\sigma} - \frac{1}{2K_\rho} \right] g_{\rm{SC}}.
\end{equation}

Continuing to second order in the spin density wave term, we find the real time partition function takes the form
\begin{equation}
	Z_2 = -\frac{g_{\rm{S}}^2}{8(2 \pi a)^4} \biggl( \frac{L}{2 \pi a} \biggr)^{-4/K_\sigma} \int \, dx_1 \, dx_2 \, dt_1 \, dt_2 :\cos [2\sqrt{2} \theta_\sigma (x_1,t_1)]: : \cos [2\sqrt{2} \theta_\sigma (x_2,t_2)]:.
\end{equation}
We then normal order the two individually normal ordered cosines, and keep only the part which gives us a new scaling compared to the first-order term to get
\begin{equation}
	Z_2 = -\frac{g_{\rm{S}}^2 a^{4/K_\sigma}}{8(2 \pi a)^4} \int \, dx_1 \, dx_2 \, dt_1 \, dt_2 \,  [m_\sigma^+ m_\sigma^-]^{-2/K_\sigma} :\cos [2\sqrt{2} \theta_\sigma (x_1,t_1)]  \cos [2\sqrt{2} \theta_\sigma (x_2,t_2)]:,
\end{equation}
where we have defined $m_\sigma^\pm = a-iv_\sigma(t_2-t_1) \pm i(x_2-x_1)$. Because they are now under normal ordering, we can expand the cosines for small $x_2-x_1$, to get
\begin{equation}
	Z_2 =-\frac{g_{\rm{S}}^2 a^{4/K_\sigma}}{(2 \pi a)^4} \int \, dx_1 \, dx_2 \, dt_1 \, dt_2 \,  [m_\sigma^+ m_\sigma^-]^{-2/K_\sigma} (x_2-x_1)^2 :(\partial_x \theta_\sigma)^2:.
\end{equation}
This is a renormalization of the coefficient of $(\partial_x \theta_\sigma)^2$, that is to say a renormalization of $K_\sigma$.
We now re-express this as a term in the first-order partition function of the operator $(\partial_x \theta_\sigma)^2$; we therefore shift to centre of mass $\tx = x_1 + x_2$ and relative $x = x_2-x_1$ coordinates (same for $t$), and leave the integration over the centre of mass coordinates alone. In order to calculate the coefficient of the term $(\partial_x \theta_\sigma)^2$, we do the RG step $a\rightarrow a-da = a(1+d\ell)$ inside the integral only, which gives us an integral form for the coefficient $\kappa_1$
\begin{equation}
	\kappa_1 = \int_{-\infty}^\infty \, dx \int_{-\infty}^0 \, dt \frac{x^2 ((1-iv_\sigma t)^2 + x^2 )^{-2/K_\sigma}}{(2\pi)^4 v_\sigma}
\end{equation}
which turns up in the renormalization of $K_\sigma$, $dK_\sigma = \kappa_1 g_{\rm{S}}^2 d\ell$. We can compute this integral exactly, to get
\begin{equation}
	\kappa_1 = \frac{1}{(2\pi)^4 v_\sigma} \frac{ \sqrt{\pi} \Gamma [2/K_\sigma - 3/2]}{2 \Gamma [2/K_\sigma]}.
\end{equation}

This term has a weak, linear dependence on $K_\sigma$, so only allowing for small changes of $K_\sigma$ we may treat it as approximately constant.


Similar calculations for the terms in $g_{\rm{U}}$ and $g_{\rm{SC}}$ give us the coupled RG equations below
\begin{eqnarray}
	\frac{ d K_\sigma}{d \ell} &= \kappa_1 g_{\rm{S}}^2 - \kappa_2 g_{\rm{U}}^2 K_\sigma^2 + \kappa_3 g_{\rm{SC}}^2, \\
	\frac{ d K_\rho}{d \ell} &=  - \kappa_2 g_{\rm{U}}^2 K_\rho^2+\kappa_3 g_{\rm{SC}}^2.
\end{eqnarray}
$\kappa_2$ is given by the integral
\begin{equation}
	\kappa_2 = \frac{4 a^{4(K_\rho+K_\sigma)+1}}{(2 \pi a)^4} \int_{-\infty}^0 \, dt \, 	\int_{-\infty}^{\infty} dx \, x^2\partial_a  \biggl\{ [(a-iv_\rho t)^2 + x^2]^{-2 K_\rho} [(a-iv_\sigma t)^2 + x^2]^{-2 K_\sigma} \biggr\},
\end{equation}
and $\kappa_3$ by
\begin{equation}
	\kappa_3 = \frac{2 a^{\left(\frac{1}{K_\rho}+\frac{1}{K_\sigma}\right)+1}}{(2 \pi a)^4} \int_{-\infty}^0 \, dt \, 	\int_{-\infty}^{\infty} dx \, x^2  \partial_a  \biggl\{ [(a-iv_\rho t)^2 + x^2]^{-\frac{1}{2K_\rho}} [(a-iv_\sigma t)^2 + x^2]^{-\frac{1}{2K_\sigma}} \biggr\}.
\end{equation}

Note that the use of the full, second order RG equations does not change qualitatively the result given in the main text using only the first order RG equations, as the second order terms only result in small changes in $K_\rho$ and $K_\sigma$.


\begin{thebibliography}{50}
	\expandafter\ifx\csname natexlab\endcsname\relax\def\natexlab#1{#1}\fi
	\expandafter\ifx\csname bibnamefont\endcsname\relax
	\def\bibnamefont#1{#1}\fi
	\expandafter\ifx\csname bibfnamefont\endcsname\relax
	\def\bibfnamefont#1{#1}\fi
	\expandafter\ifx\csname citenamefont\endcsname\relax
	\def\citenamefont#1{#1}\fi
	\expandafter\ifx\csname url\endcsname\relax
	\def\url#1{\texttt{#1}}\fi
	\expandafter\ifx\csname urlprefix\endcsname\relax\def\urlprefix{URL }\fi
	\providecommand{\bibinfo}[2]{#2}
	\providecommand{\eprint}[2][]{\url{#2}}
	
	\bibitem{Nayak+2008}
	Chetan Nayak, Steven~H. Simon, Ady Stern, Michael Freedman, and Sankar
	Das~Sarma.
	\newblock Non-abelian anyons and topological quantum computation.
	\newblock {\em Rev. Mod. Phys.}, 80:1083--1159, Sep 2008.
	
	\bibitem{leijnse12}
	Martin Leijnse and Karsten Flensberg.
	\newblock Introduction to topological superconductivity and {M}ajorana
	fermions.
	\newblock {\em Semicond. Sci. Technol.}, 27(12):124003, December 2012.
	
	\bibitem{Alicea+2012}
	Jason Alicea.
	\newblock New directions in the pursuit of majorana fermions in solid state
	systems.
	\newblock {\em Rep. Prog. Phys.}, 75(7):076501, 2012.
	
	\bibitem{beenakker13}
	{C.W.J.} Beenakker.
	\newblock Search for {M}ajorana fermions in superconductors.
	\newblock {\em Ann. Rev. Cond. Mat. Phys.}, 4(1):113, 2013.
	
	\bibitem{majorana37}
	E.~Majorana.
	\newblock Symmetrical theory of electrons and positrons.
	\newblock {\em Nuovo Cimento}, 14:171, 1937.
	
	\bibitem{read00}
	N.~Read and Dmitry Green.
	\newblock Paired states of fermions in two dimensions with breaking of parity
	and time-reversal symmetries and the fractional quantum {H}all effect.
	\newblock {\em Phys. Rev. B}, 61:10267, 2000.
	
	\bibitem{Ivanov+2001}
	D.~A. Ivanov.
	\newblock Non-abelian statistics of half-quantum vortices in $\mathit{p}$-wave
	superconductors.
	\newblock {\em Phys. Rev. Lett.}, 86:268--271, Jan 2001.
	
	\bibitem{Kitaev+2001}
	A~Yu Kitaev.
	\newblock Unpaired majorana fermions in quantum wires.
	\newblock {\em Physics-Uspekhi}, 44(10S):131, 2001.
	
	\bibitem{Fu+2008}
	Liang Fu and C.~L. Kane.
	\newblock Superconducting proximity effect and majorana fermions at the surface
	of a topological insulator.
	\newblock {\em Phys. Rev. Lett.}, 100:096407, Mar 2008.
	
	\bibitem{Fu+2009}
	Liang Fu and C.~L. Kane.
	\newblock Josephson current and noise at a
	superconductor/quantum-spin-hall-insulator/superconductor junction.
	\newblock {\em Phys. Rev. B}, 79:161408, Apr 2009.
	
	\bibitem{Sato+2009}
	Masatoshi Sato and Satoshi Fujimoto.
	\newblock Topological phases of noncentrosymmetric superconductors: Edge
	states, majorana fermions, and non-abelian statistics.
	\newblock {\em Phys. Rev. B}, 79:094504, Mar 2009.
	
	\bibitem{Sato+2009B}
	Masatoshi Sato, Yoshiro Takahashi, and Satoshi Fujimoto.
	\newblock Non-abelian topological order in $s$-wave superfluids of ultracold
	fermionic atoms.
	\newblock {\em Phys. Rev. Lett.}, 103:020401, Jul 2009.
	
	\bibitem{Oreg+2010}
	Yuval Oreg, Gil Refael, and Felix von Oppen.
	\newblock Helical liquids and majorana bound states in quantum wires.
	\newblock {\em Phys. Rev. Lett.}, 105:177002, Oct 2010.
	
	\bibitem{Lutchyn+2010}
	Roman~M. Lutchyn, Jay~D. Sau, and S.~Das~Sarma.
	\newblock Majorana fermions and a topological phase transition in
	semiconductor-superconductor heterostructures.
	\newblock {\em Phys. Rev. Lett.}, 105:077001, Aug 2010.
	
	\bibitem{fendley12}
	Paul Fendley.
	\newblock Parafermionic edge zero modes in {$Z_n$}-invariant spin chains.
	\newblock {\em J. Stat. Mech.}, 2012(11):P11020, 2012.
	
	\bibitem{Lindner+2012}
	Netanel~H. Lindner, Erez Berg, Gil Refael, and Ady Stern.
	\newblock Fractionalizing majorana fermions: Non-abelian statistics on the
	edges of abelian quantum hall states.
	\newblock {\em Phys. Rev. X}, 2:041002, Oct 2012.
	
	\bibitem{Clarke+2013}
	David~J Clarke, Jason Alicea, and Kirill Shtengel.
	\newblock Exotic non-abelian anyons from conventional fractional quantum hall
	states.
	\newblock {\em Nat. Comm.}, 4:1348, 2013.
	
	\bibitem{alicea15}
	Jason Alicea and Paul Fendley.
	\newblock Topological phases with parafermions: theory and blueprints.
	\newblock arXiv:1504.02476 [cond-mat.str-el], 2015.
	
	\bibitem{Deng+2012}
	M.~T. Deng, C.~L. Yu, G.~Y. Huang, M.~Larsson, P.~Caroff, and H.~Q. Xu.
	\newblock Anomalous zero-bias conductance peak in a nb–insb nanowire–nb
	hybrid device.
	\newblock {\em Nano Letters}, 12(12):6414--6419, 2012.
	
	\bibitem{Mourik+2012}
	V.~Mourik, K.~Zuo, S.~M. Frolov, S.~R. Plissard, E.~P. A.~M. Bakkers, and L.~P.
	Kouwenhoven.
	\newblock Signatures of majorana fermions in hybrid
	superconductor-semiconductor nanowire devices.
	\newblock {\em Science}, 336(6084):1003--1007, 2012.
	
	\bibitem{Wesperen+2013}
	Ilse van Weperen, Sébastien~R. Plissard, Erik P. A.~M. Bakkers, Sergey~M.
	Frolov, and Leo~P. Kouwenhoven.
	\newblock Quantized conductance in an insb nanowire.
	\newblock {\em Nano Letters}, 13(2):387--391, 2013.
	
	\bibitem{Nadj+2014}
	Stevan Nadj-Perge, Ilya~K Drozdov, Jian Li, Hua Chen, Sangjun Jeon, Jungpil
	Seo, Allan~H MacDonald, B~Andrei Bernevig, and Ali Yazdani.
	\newblock Observation of majorana fermions in ferromagnetic atomic chains on a
	superconductor.
	\newblock {\em Science}, 346(6209):602--607, 2014.
	
	\bibitem{rokhinson12}
	Leonid~P. Rokhinson, Xinyu Liu, and Jacek~K. Furdyna.
	\newblock The fractional a.c. {J}osephson effect in a
	semiconductor-superconductor nanowire as a signature of {M}ajorana particles.
	\newblock {\em Nat. Phys.}, 8:795, 2012.
	
	\bibitem{Maier+2015}
	L.~{Maier}, E.~{Bocquillon}, M.~{Grimm}, J.~B. {Oostinga}, C.~{Ames},
	C.~{Gould}, C.~{Br{\"u}ne}, H.~{Buhmann}, and L.~W. {Molenkamp}.
	\newblock Phase-sensitive squids based on the 3d topological insulator hgte.
	\newblock {\em Physica Scripta Volume T}, 164(1):014002, December 2015.
	
	\bibitem{Wiedenmann+2015}
	J.~{Wiedenmann}, E.~{Bocquillon}, R.~S. {Deacon}, S.~{Hartinger},
	O.~{Herrmann}, T.~M. {Klapwijk}, L.~{Maier}, C.~{Ames}, C.~{Br{\"u}ne},
	C.~{Gould}, A.~{Oiwa}, K.~{Ishibashi}, S.~{Tarucha}, H.~{Buhmann}, and L.~W.
	{Molenkamp}.
	\newblock 4$\pi$-periodic josephson supercurrent in hgte-based topological
	josephson junctions.
	\newblock {\em Nat. Comm.}, 7(10303), January 2016.
	
	\bibitem{Zhang+2014}
	Fan Zhang and C.~L. Kane.
	\newblock Time-reversal-invariant ${Z}_{4}$ fractional josephson effect.
	\newblock {\em Phys. Rev. Lett.}, 113:036401, Jul 2014.
	
	\bibitem{klinovaja13b}
	Jelena Klinovaja and Daniel Loss.
	\newblock Time-reversal invariant parafermions in interacting {R}ashba
	nanowires.
	\newblock {\em Phys. Rev. B}, 90:045118, Jul 2014.
	
	\bibitem{Klinovaja+2015}
	Jelena Klinovaja and Daniel Loss.
	\newblock Fractional charge and spin states in topological insulator
	constrictions.
	\newblock {\em Phys. Rev. B}, 92:121410, Sep 2015.
	
	\bibitem{Orth+2015}
	Christoph~P. Orth, Rakesh~P. Tiwari, Tobias Meng, and Thomas~L. Schmidt.
	\newblock Non-abelian parafermions in time-reversal-invariant interacting
	helical systems.
	\newblock {\em Phys. Rev. B}, 91:081406(R), Feb 2015.
	
	\bibitem{Haim+2014}
	Arbel Haim, Anna Keselman, Erez Berg, and Yuval Oreg.
	\newblock Time-reversal-invariant topological superconductivity induced by
	repulsive interactions in quantum wires.
	\newblock {\em Phys. Rev. B}, 89:220504, Jun 2014.
	
	\bibitem{winkler_book}
	Roland Winkler.
	\newblock {\em Spin-Orbit Coupling Effects in Two-Dimensional Electron and Hole
		Systems}, volume 191 of {\em Springer Tracts in Modern Physics}.
	\newblock Springer, 2003.
	
	\bibitem{moroz00}
	A.~V. Moroz, K.~V. Samokhin, and C.~H.~W. Barnes.
	\newblock Theory of quasi-one-dimensional electron liquids with spin-orbit
	coupling.
	\newblock {\em Phys. Rev. B}, 62:16900, 2000.
	
	\bibitem{moroz00a}
	A.~V. Moroz, K.~V. Samokhin, and C.~H.~W. Barnes.
	\newblock Spin-orbit coupling in interacting quasi-one-dimensional electron
	systems.
	\newblock {\em Phys. Rev. Lett.}, 84:4164, May 2000.
	
	\bibitem{Starykh+2008}
	Suhas Gangadharaiah, Jianmin Sun, and Oleg~A. Starykh.
	\newblock Spin-orbital effects in magnetized quantum wires and spin chains.
	\newblock {\em Phys. Rev. B}, 78:054436, Aug 2008.
	
	\bibitem{kainaris15}
	Nikolaos Kainaris and Sam~T. Carr.
	\newblock Emergent topological properties in interacting one-dimensional
	systems with spin-orbit coupling.
	\newblock {\em Phys. Rev. B}, 92:035139, Jul 2015.
	
	\bibitem{cardy96}
	John Cardy.
	\newblock {\em Scaling and Renormalization in Statistical Physics}.
	\newblock Cambridge University Press, 1996.
	
	\bibitem{Oreg+2014}
	Yuval Oreg, Eran Sela, and Ady Stern.
	\newblock Fractional helical liquids in quantum wires.
	\newblock {\em Phys. Rev. B}, 89:115402, Mar 2014.
	
	\bibitem{Giamarchi+2004}
	Thierry Giamarchi.
	\newblock Quantum physics in one dimension, 2004.
	
	\bibitem{Pikulin+2012}
	D~I Pikulin, J~P Dahlhaus, M~Wimmer, H~Schomerus, and C~W~J Beenakker.
	\newblock A zero-voltage conductance peak from weak antilocalization in a
	majorana nanowire.
	\newblock {\em New J. Phys.}, 14(12):125011, 2012.
	
	\bibitem{Liu+2012}
	Jie Liu, Andrew~C. Potter, K.~T. Law, and Patrick~A. Lee.
	\newblock Zero-bias peaks in the tunneling conductance of spin-orbit-coupled
	superconducting wires with and without majorana end-states.
	\newblock {\em Phys. Rev. Lett.}, 109:267002, Dec 2012.
	
	\bibitem{Fidkowski+2011}
	Lukasz Fidkowski and Alexei Kitaev.
	\newblock Topological phases of fermions in one dimension.
	\newblock {\em Phys. Rev. B}, 83:075103, Feb 2011.
	
	\bibitem{turner11}
	Ari~M. Turner, Frank Pollmann, and Erez Berg.
	\newblock Topological phases of one-dimensional fermions: An entanglement point
	of view.
	\newblock {\em Phys. Rev. B}, 83:075102, Feb 2011.
	
	\bibitem{Sela+2011}
	Eran Sela, Alexander Altland, and Achim Rosch.
	\newblock Majorana fermions in strongly interacting helical liquids.
	\newblock {\em Phys. Rev. B}, 84:085114, Aug 2011.
	
	\bibitem{Dominguez+2012}
	Fernando Dom\'{\i}nguez, Fabian Hassler, and Gloria Platero.
	\newblock Dynamical detection of majorana fermions in current-biased nanowires.
	\newblock {\em Phys. Rev. B}, 86:140503, Oct 2012.
	
	\bibitem{Shapiro+1963}
	Sidney Shapiro.
	\newblock Josephson currents in superconducting tunneling: The effect of
	microwaves and other observations.
	\newblock {\em Phys. Rev. Lett.}, 11:80--82, Jul 1963.
	
	\bibitem{Zumbuhl+2014}
	C.~P. Scheller, T.-M. Liu, G.~Barak, A.~Yacoby, L.~N. Pfeiffer, K.~W. West, and
	D.~M. Zumb\"uhl.
	\newblock Possible evidence for helical nuclear spin order in gaas quantum
	wires.
	\newblock {\em Phys. Rev. Lett.}, 112:066801, Feb 2014.
	
	\bibitem{Hevroni+2015}
	R~Hevroni, V~Shelukhin, M~Karpovski, M~Goldstein, E~Sela, H~Shtrikman, and
	A~Palevski.
	\newblock Suppression of coulomb blockade peaks by electronic correlations in
	inas nanowires.
	\newblock {\em arXiv:1504.03463}, 2015.
	
	\bibitem{li15}
	Tingxin Li, Pengjie Wang, Hailong Fu, Lingjie Du, Kate~A. Schreiber, Xiaoyang
	Mu, Xiaoxue Liu, Gerard Sullivan, Gabor~A. Csathy, Xi~Lin, and Rui-Rui Du.
	\newblock Observation of a helical luttinger-liquid in inas/gasb quantum spin
	hall edges.
	\newblock {\em arXiv:1507.08362}, 2015.
	
\end{thebibliography}
\end{document}